\documentstyle[12pt]{article}
\begin{document} 
\global\parskip 6pt
\def\mtx#1{\quad\hbox{{#1}}\quad}
\newcommand{\be}{\begin{equation}}
\newcommand{\ee}{\end{equation}}
\newcommand{\bea}{\begin{eqnarray}}
\newcommand{\eea}{\end{eqnarray}}
\newcommand{\non}{\nonumber}

\begin{titlepage}
\begin{flushright}
SINP-TNP/01-04\\
hep-th/0102155
\end{flushright}
\vspace*{1cm}
\begin{center}
{\Large\bf Exact Results for the BTZ Black Hole}\\
\vspace*{2cm}
Danny Birmingham\footnote{Email: dannyb@pop3.ucd.ie}\\
{\em Department of Mathematical Physics,\\
University College Dublin,\\
Belfield, Dublin 4, Ireland}\\
\vspace*{.5cm}
Ivo Sachs\footnote{Email: ivo@theorie.physik.uni-muenchen.de}\\
{\em Theoretische Physik,\\
Ludwig-Maximilians Universit\"{a}t,\\
Theresienstrasse 37,\\
80333 Munich, Germany}\\
\vspace*{.5cm}
Siddhartha Sen\footnote{Email: sen@maths.tcd.ie; On Leave from: School of
Mathematics, Trinity College Dublin, Ireland}\\
{\em Saha Institute of Nuclear Physics,\\
1/AF Bidhannagar,\\
Calcutta 700 064, India}\\
\vspace{1cm}
\begin{abstract}
In this review, we summarize exact results for the three-dimensional
BTZ black hole. We use rigorous mathematical results
to clarify
the general structure and properties
of this black hole spacetime and its microscopic description. 
In particular, we study the formation of the black hole by point particle
collisions, leading to an exact analytic determination of the Choptuik
scaling parameter. 
We also show that a `No Hair Theorem' follows immediately
from a mathematical theorem of hyperbolic geometry, due to Sullivan.
A microscopic understanding of the Bekenstein-Hawking
entropy, and decay rate for massless scalars, is shown to follow
from standard results of conformal field theory.

\end{abstract}
\vspace{1cm}
February 2001
\end{center}
\end{titlepage} 

\section{Introduction}
The challenge of understanding the microscopic properties 
of black hole physics is an important step in the quest for a 
consistent quantum theory of gravity. Recent developments in string 
theory have led to significant advances
in our understanding of the microscopic structure of black holes.
The microscopic derivation of the
Bekenstein-Hawking entropy of a string theory black hole
is perhaps the most notable \cite{SV}. This result led to a more
thorough investigation of the nature of black holes within
the context of string theory; for recent reviews see,
for example, \cite{Mal1}-\cite{Peet2}.
It became clear that many aspects of a large class of
extremal and near-extremal black holes can be given a conformal
field theory description. The black holes in question are 
solutions of the low-energy equations of motion of superstring
theory. One of the most significant aspect of these black holes
is the fact that they are charged with respect to certain 
RR-fields. These RR-charges are in turn carried by D-branes 
in string theory \cite{Pol}.
Exact agreement was found between the
Bekenstein-Hawking
entropy of a classical five-dimensional extremal RR-charged
black hole and the entropy of states of a corresponding
D-brane system with the same quantum numbers \cite{SV}.

On the other hand it is instructive to have at one's disposal
a lower-dimensional toy model, as this often affords the possibility
of a more detailed analysis.
The construction of a black hole spacetime in $(2+1)$
dimensions, with negative cosmological constant, provides
one such example. This is known as the Ba\~{n}ados-Teitelboim-Zanelli
(BTZ) black hole \cite{BTZ1}, and it
warrants study in its own right; for a review, see \cite{Carlip1}.
A key feature of this model lies in the simplicity of
its construction; it is a constant negatively curved spacetime,
and it is obtained as a discrete quotient of three-dimensional
anti-de Sitter space \cite{BTZ2}. In particular, there is no curvature 
singularity. Even so, all characteristic features such as the event horizon 
and Hawking radiation are present so that this model is a 
genuine black hole. Furthermore, despite its simplicity, the BTZ black hole 
plays a significant role in many of the
recent developments in string theory
\cite{Skend}, \cite{hyun}-\cite{Strom}.
In particular, a generic feature of all string theory black holes for 
which an exact counting of microstates is possible, is that the near
horizon geometry of these solutions is that of a BTZ black hole. 
Our aim in this review is to establish a number of exact
results for the kinematical and dynamical properties of
the BTZ black hole. The approach presented here is complementary 
to the semi-classical quantization which focuses on quantum
field theory in a black hole background. Here, we are interested
in the microscopic properties of the black hole itself. We obtain 
these results without referring to any 
specific underlying microscopic theory for gravity such as string theory. 
While these results shed light on the general structure
of the $(2+1)$-dimensional
spacetime itself, they also  help to understand universal
properties of certain higher-dimensional string theory black holes. For a 
discussion of string theory black holes and their connection to the 
BTZ-black hole see for example, \cite{Peet1}-\cite{Peet2}.

Firstly, we focus on the formation of the black
hole from point particle collisions \cite{Mats1}-\cite{BS1}. We show that
formation can be understood at a purely algebraic level,
in terms of isometries of anti-de Sitter space. This leads
to an exact analytic understanding of the associated Choptuik scaling
parameter, and represents the first such exact determination \cite{BS1}.
Interest in black hole formation within the context
of numerical relativity has increased due to the
critical scaling behaviour discovered by Choptuik \cite{Chop}.
It was observed numerically that the threshold for black
hole formation has a simple structure in the space of initial
data. In particular, the black hole mass parameter, for example,
exhibits a certain universal power-law
scaling behaviour; for a review, see \cite{Gund}. It is important to have
a model where this behaviour can be studied exactly and analytically,
and the BTZ black hole provides such an example \cite{BS1}.

Following on from this, we invoke a precise mathematical theorem
of hyperbolic geometry, due to Sullivan \cite{Sull,McM},
to establish a `No Hair Theorem' for the
BTZ black hole \cite{BKSW}. This result shows that the BTZ black hole
can be parametrized by at most two parameters, its mass
and angular momentum.
Furthermore, the theorem of Sullivan provides a precise notion
of holography, whereby the three-dimensional spacetime structure
is completely determined
in terms of certain two-dimensional boundary data.
The relevance of holography in gravitational systems
has often been discussed \cite{thooft,Suss}.
Again, this concept has received
exciting impetus from recent developments in string theory.
In particular, the AdS/CFT conjecture of Maldacena
\cite{Malda1}-\cite{Witten2}
states
that string theory defined on an anti-de Sitter
background is dual, in a particular limit, to superconformal
field theory defined on the boundary of anti-de Sitter space;
for a review, see \cite{Malda2}. 
This holographic conjecture has received enormous
attention recently, and has led to important advances
in our understanding of the dynamics of gauge theory and gravity.
However, while the dynamical conjecture of Maldacena remains to be proved,
it is satisfying to have an exact mathematical theorem
which establishes a precise notion of holography in a kinematical
sense. The fact that the BTZ black hole satisfies the requirements of this
theorem again shows the relevance of this model within the
wider context of string theory.
      
As mentioned above, the microscopic derivation of the entropy of
string theory black holes was carried out
by counting the excitations of certain
D-brane configurations at weak coupling where spacetime is flat. 
Supersymmetry then relates this configuration to the corresponding
black hole spacetime with the given RR-charges. An interesting question is 
whether it is possible to count the microstates without recourse to a
specific microscopic realisation in terms of D-branes. 
It turns out that this is indeed possible due to an important
result concerning the asymptotic symmetry algebra
of the BTZ black hole. It was shown by Brown and Henneaux \cite{BH}
that the asymptotic symmetric algebra
consists of a Virasoro algebra with both left-moving and right-moving
sectors. A formula due to Cardy \cite{Cardy} for the degeneracy of states 
in a conformal field theory 
can then be used to give a universal microscopic derivation of the BTZ entropy
\cite{Strom,BSS2}. Furthermore, this understanding
of BTZ entropy lies at the heart
of many of the derivations for higher-dimensional
string theory black holes \cite{Sfetsos,Strom}, \cite{BL}-\cite{CL2}.
This is due to the fact that these black holes
have a near-horizon geometry containing
the BTZ black hole. 
We also show that an exact convergent expansion
for the degeneracy of states of a
conformal field theory can be used
to compute the corrections to the Bekenstein-Hawking entropy \cite{BS2}.
This expansion, due to Rademacher \cite{Rade,DMMV}, is a remarkable
result from analytic number theory
and generalizes the Cardy formula mentioned above. Due to its exact
nature, the expansion gives a precise determination
of the form of the corrections to the Bekenstein-Hawking
entropy formula \cite{BS2,Carlip2}. An alternative way of counting the BTZ 
microstates was previously suggested by Carlip \cite{Carlip1b}.

Having discussed the formation process, and
resulting kinematical properties, it remains to understand the decay of
non-extremal black holes. This involves studying the absorption
of quanta by the black hole,  and then 
allowing it to evaporate, via Hawking radiation, back to extremality. In 
\cite{Callan}-\cite{GK}, the low energy
scattering cross sections and decay rates for a massless 
minimally coupled scalar field were computed for a large class of 
four- and five-dimensional black holes, 
and agreement was found with conformal field theory or effective string
theory predictions. In each of these cases, the result 
relied on a particular matching
of solutions between a near-horizon region and an asymptotic
region.
For certain ranges of parameters inherent to the problem, this matching 
agrees with a conformal field theory description. 

Here, we study the propagation a massless minimally coupled scalar
field in the background geometry
of the BTZ black hole. 
The special feature is that the wave equation can be solved exactly,
without any approximations \cite{Lar,IS}.
This allows us to determine exactly the range of energy and 
angular momentum of the scattered field, for which the the 
decay rate agrees with a conformal field
theory description \cite{BSS1}. We find
agreement for energies small in comparison to the size of the black
hole, and to the curvature scale of the spacetime; in addition, one
is restricted to  the zero angular momentum wave. In this region, 
however, agreement is found for all values of mass
and angular momentum, and thus the conformal field theory
description is not restricted to a near-extremal limit.  Finally, we
provide a microscopic derivation of the Hawking decay rate \cite{ES}.
This is achieved by first showing how the
conformal field theory which represents the asymptotic isometries
of the BTZ black hole is perturbed by the presence of matter fields.
The transition probabilities in the black hole which are induced by
this perturbation are then calculated, and we
find agreement with the semiclassical
Hawking decay rate.

The plan of this article is as follows.
In section 2, we present the basic construction and properties
of the BTZ black hole. In section 3, we discuss the connection
between the Gott time machine and the formation
of the BTZ black hole from point particle collisions. This yields
an exact analytic determination
of the Choptuik scaling parameter.
In section 4, we show how Sullivan's theorem
of hyperbolic geometry provides a `No Hair Theorem' for
the BTZ black hole, and we also establish the holographic nature
of the spacetime. This is followed by the
microscopic derivation of the entropy from the Brown-Henneaux
algebra.
We also use a convergent Rademacher expansion to compute
corrections to the Bekenstein-Hawking entropy in an exact way.
In section 5, we study the low energy dynamics of the BTZ black hole.
In particular, we compute the exact decay rate for massless minimally
coupled scalars, and show how it can be interpreted in terms
of a left-right symmetric conformal field theory.
We present our conclusions in section 6.

\section{Construction and Properties}
We begin by recalling the metric for the BTZ black hole. In terms
of Schwarzschild
coordinates, the line element is given by \cite{BTZ1,Carlip1}
\bea
ds^{2} = - \left(-M + \frac{r^{2}}{l^{2}} +
\frac{J^{2}}{4r^{2}}\right)
dt^{2} + \left(-M + \frac{r^{2}}{l^{2}} +
\frac{J^{2}}{4r^{2}}\right)^{-1}
dr^{2} + r^{2} \left( d\phi - \frac{J}{2r^{2}}dt\right)^{2},
\label{btzmet}
\eea
where
$M$ and $J$ are the mass and angular momentum parameters of the black
hole.
This metric satisfies the vacuum Einstein equations
in $2+1$ dimensions, with a negative cosmological constant
$\Lambda = -1/l^{2}$, namely
\bea
R_{\mu\nu} - \frac{1}{2}g_{\mu\nu}R = \frac{1}{l^{2}}g_{\mu\nu}.
\label{einstein}
\eea
The metric is singular at the location of the inner
and outer horizons $r = r_{\pm}$, defined by
\be\label{rpm}
r_{\pm}^{2} = \frac{Ml^{2}}{2}\left(1 \pm \sqrt{1 -
\frac{J^{2}}{M^{2}l^{2}}}\;\right).
\label{horizons}
\ee
Thus, we can express the mass and angular momentum in terms
of $r_{\pm}$ as
\bea
M = \frac{r_{+}^{2} + r_{-}^{2}}{l^{2}},\;\; J = \frac{2 r_{+}r_{-}}{l}.
\label{mass}
\eea

We shall use units for which Newton's constant
satisfies $8G = 1$.
The $M=-1$, $J=0$ metric is then recognized as that of three-dimensional
anti-de Sitter space ($AdS_{3}$). It is separated by a mass gap
from the $M=0$,
$J=0$ ``massless black hole."

For later use,
let us recall some elements of the canonical
approach to the BTZ black hole \cite{BTZ2}. For this,
we consider a general stationary, axially symmetric,
configuration parametrized by
\be\label{asa}
ds^2=-(N^\perp(r))^2dt^2+\frac{dr^2}{f^2(r)}
+r^2\left(N^\phi(r)dt+d\phi\right)^2\ ,
\qquad \phi\in[0,2\pi)\ ,
\ee
where $N^\perp$ and $N^\phi$ are the usual lapse and shift functions.
Upon solving the constraint equations, one finds that
\bea
N^\perp(r)&=&f(r)N(\infty),\non\\
N^\phi(r)&=&-\frac{J}{2r^2}N(\infty)+N^\phi(\infty),\\
f^2(r)&=&\frac{r^2}{l^2}-M+\frac{J^2}{4r^2}\ .\non
\eea
Thus far, $N(\infty),N^\phi(\infty),M$ and $J$ are merely 
integration constants. On the other hand, if we now consider
arbitrary variations $\delta g_{\mu\nu}$ approaching the form
(\ref{asa}) for large $r$,  it is easy to see
that under such variations the
action picks up a boundary contribution 
\be
\delta S=(t_2-t_1)\left[N(\infty)\delta M-N^\phi(\infty)\delta 
J\right]+\delta {\cal{B}}\ ,
\ee
where ${\cal{B}}$ parametrizes the possibility of adding boundary terms to the 
Einstein-Hilbert action. In order for the variation of the action to 
vanish for the BTZ solution, we must choose
\be\label{boun}
{\cal{B}}=(t_2-t_1)\left[-N(\infty)M+N^\phi(\infty)J\right]\ .
\ee
We now see that $M$ and $J$ appear as variables conjugate to the lapse
and
shift functions,
justifying their interpretation as mass and angular momentum,
respectively. 

It is straightforward to check that
any solution of the vacuum Einstein
equations (\ref{einstein}) is also a space of constant negative curvature.
Thus, the BTZ black hole is locally
isometric to $AdS_{3}$.
Moreover, it is known that the BTZ black hole is obtained
by performing a quotient of $AdS_{3}$ by a discrete
finitely generated group of isometries of $AdS_{3}$.
For completeness, we include the relevant parts of this construction,
as it plays a crucial role in the remaining sections.

A standard representation \cite{Carlip1} of $AdS_{3}$ may be obtained from
a flat spacetime ${\mathbf R}^{2,2}$, with coordinates
$(X_{1}, X_{2}, T_{1}, T_{2})$, and metric
\bea
ds^{2} = dX_{1}^{2} + dX_{2}^{2} - dT_{1}^{2} - dT_{2}^{2}.
\eea
The induced metric on the submanifold
\bea
X_{1}^{2} + X_{2}^{2} - T_{1}^{2} - T_{2}^{2} = -l^{2},
\label{ads}
\eea
then corresponds to the $AdS_{3}$ metric. From (\ref{ads}), it is clear
that the isometry group of $AdS_{3}$ is
$SO(2,2)$. Equivalently, we can combine the
coordinates
into an $SL(2,{\mathbf R})$ matrix
\bea
X =
\frac{1}{l}\left (
\begin{array}{cc}
T_{1} + X_{1}&T_{2} + X_{2}\\
-T_{2} + X_{2}& T_{1} - X_{1}
\end{array}
\right ),
\eea
with $\mathrm{det}\;X = 1$.
Hence, $AdS_{3}$ can be viewed as the group manifold of
$SL(2, {\mathbf R})$, with  the isometries being represented
by left and right multiplication via
$X \rightarrow \rho_{L} X \rho_{R}$. Here,
$\rho_{L}$ and $\rho_{R}$ are elements of $SL(2,{\mathbf R})$.
One has a further ${\mathbf Z}_{2}$ identification
of isometries, namely
$(\rho_{L}, \rho_{R}) \sim (-\rho_{L}, -\rho_{R})$.
This follows from the fact that
$SO(2,2) \approx SL(2,{\mathbf R})
\times SL(2, {\mathbf R})/{\mathbf Z}_{2}$.

The key point in the construction of the BTZ
black hole is to recognize that the requirement
of periodicity in the angular coordinate $\phi$ is implemented
by identifying points of anti-de Sitter space by the isometry
$(\rho_{L}, \rho_{R})$, where
\bea
\rho_{L} =
\left (
\begin{array}{cc}
e^{\pi(r_{+} - r_{-})/l}&0\\
0& e^{-\pi(r_{+} - r_{-})/l}
\end{array}
\right ), \;\;
\rho_{R} =
\left (
\begin{array}{cc}
e^{\pi(r_{+} + r_{-})/l}&0\\
0& e^{-\pi(r_{+} + r_{-})/l}
\end{array}
\right ).
\eea
The BTZ black hole is then given \cite{Carlip1,BTZ2}
as the quotient space
$AdS_{3}/\langle(\rho_{L},\rho_{R})\rangle$,
where $\langle(\rho_{L},\rho_{R})\rangle$
denotes the group generated by the isometry $(\rho_{L}, \rho_{R})$.

We shall also require certain properties of the Euclidean section of
the black hole \cite{Carlip1,Carlip3}.
The metric is obtained from (\ref{btzmet})
by performing the continuation $t = i\tau_{E}$ and $J = -iJ_{E}$,
leading to the line element
\bea
ds_{E}^{2} = \left(-M + \frac{r^{2}}{l^{2}} -
\frac{J_{E}^{2}}{4r^{2}}\right)
d\tau_{E}^{2} + \left(-M + \frac{r^{2}}{l^{2}} -
\frac{J_{E}^{2}}{4r^{2}}\right)^{-1}
dr^{2} + r^{2} \left( d\phi - \frac{J_{E}}{2r^{2}}d\tau_{E}\right)^{2}.
\eea
The parameters $r_{\pm}$ are now given by
\be
r_{+} = \left[\frac{Ml^{2}}{2}\left(1 + \sqrt{1 +
\frac{J_{E}^{2}}{M^{2}l^{2}}}\;\right)\right]^{1/2},\;\;
r_{-} = \left[\frac{Ml^{2}}{2}\left(1 - \sqrt{1 +
\frac{J_{E}^{2}}{M^{2}l^{2}}}\;\right)\right]^{1/2} \equiv -i\mid\!
r_{-}\!\mid .
\ee

Three-dimensional hyperbolic space (a space of constant
negative curvature with Euclidean signature) is denoted by $H^{3}$.
The Euclidean BTZ black hole is obtained as a
quotient of $H^{3}$ by a discrete finitely generated
group of isometries of $H^{3}$.
The precise form of the identifications can be seen
by writing the metric for $H^{3}$
in the upper-half space coordinates $(x,y,z)$ defined by
\cite{Carlip1,Carlip3}
\bea
x &=& \left(\frac{r^{2} - r_{+}^{2}}{r^{2} - r_{-}^{2}}\right)^{1/2}\;
\cos\left( \frac{r_{+}}{l^{2}}\tau_{E} + \frac{\mid\! r_{-}\!
\mid}{l}\phi\right)\; \exp\left(\frac{r_{+}}{l}\phi
- \frac{\mid\! r_{-}
\!\mid}{l^{2}} \tau_{E}\right),\non\\
y &=& \left(\frac{r^{2} - r_{+}^{2}}{r^{2} - r_{-}^{2}}\right)^{1/2}\;
\sin\left( \frac{r_{+}}{l^{2}}\tau_{E} + \frac{\mid\! r_{-}\!
\mid}{l}\phi\right)\; \exp\left(\frac{r_{+}}{l}\phi - \frac{\mid\! r_{-}\!
\mid}{l^{2}} \tau_{E}\right),\non\\
z &=& \left(\frac{r_{+}^{2} - r_{-}^{2}}{r^{2} - r_{-}^{2}}\right)^{1/2}\;
\exp\left(\frac{r_{+}}{l}\phi - \frac{\mid\! r_{-}\!
\mid}{l^{2}} \tau_{E}\right).
\label{uhs}
\eea
The metric becomes
\bea\label{up}
ds_{E}^{2} = \frac{l^{2}}{z^{2}}(dx^{2} + dy^{2} + dz^{2}),\;\;
z>0.
\eea
It follows that the BTZ metric is locally isometric to $H^{3}$.
However, from (\ref{uhs}), we see that periodicity in the Schwarzschild
coordinate $\phi$ is implemented \cite{Carlip1,Carlip3}
via the identifications
\bea
& &(x,y,z) \sim \label{Gamma}\\
& & e^{2\pi r_{+}/l}\left(
x  \cos\left(\frac{2\pi \mid\! r_{-}\!\mid}{l}\right)
-y \sin\left(\frac{2\pi \mid\! r_{-}\!\mid}{l}\right),
x \sin\left(\frac{2\pi\mid\! r_{-}\!\mid}{l}\right)
+y \cos\left(\frac{2\pi \mid\! r_{-}\!\mid}{l}\right),
z\right).\non
\eea

To determine the topology of the resulting Euclidean black hole, we
introduce spherical coordinates on
the upper-half space defined by \cite{Carlip1,Carlip3}
\bea
(x,y,z) = (R \cos \theta \cos \chi, R \sin \theta \cos \chi,
R \sin \chi),
\eea
with $\theta \in[0,2\pi], \chi \in[0,\pi/2]$.
The line element is then written as
\bea
ds_{E}^{2} = \frac{l^{2}}{\sin^{2}\chi}\left( \frac{dR^{2}}{R^{2}}
+ d\chi^{2} + \cos^{2}\chi \;d\theta^{2}\right),
\eea
and the  identifications (\ref{Gamma}) become
\bea
(R,\theta,\chi) \sim
\left(R\;e^{2 \pi r_{+}/l}, \theta + \frac{2 \pi \mid\!
r_{-}\!\mid}{l}, \chi\right).
\eea
A fundamental region is the space between the hemispheres $R=1$ and
$R= e^{2\pi r_{+}/l}$ with the inner and outer boundaries
identified along a radial line, followed by a $2\pi r_{+}/l$ rotation
about the $z$-axis. Topologically, the resulting manifold is
a solid torus \cite{Carlip1,Carlip3}. For $\chi \neq \pi/2$,
each slice of fixed $\chi$ is a $2$-torus with periodic coordinates
$\ln R$ and $\theta$.

Finally, we recall that the Hawking temperature $T_{H}$,
the length of the event horizon $A_{H}$, and the angular
velocity at the event horizon $\Omega_{H}$,
are given by \cite{Carlip1}
\bea
T_{H} = \frac{r_{+}^{2} - r_{-}^{2}}{2 \pi \ell^{2}r_{+}},\;\;
A_{H} = 2 \pi r_{+},\;\; \Omega_{H} = \frac{J}{2 r_{+}^{2}}.
\label{temp}
\eea

\section{Formation and Choptuik Scaling}

Within the context of numerical relativity, one of the most significant
recent results is the evidence for Choptuik scaling in
the formation of black hole spacetimes \cite{Chop}.
In particular, the threshold for black hole formation
in the space of initial data was observed to have a
surprisingly simple structure. Specifically, one considers
a generic smooth one-parameter
family of initial data (labelled by $p$),
such that for large values of the
parameter $p$ a black hole is formed, while no black hole
is formed for small $p$. The mass $M$ of the black hole then satisfies
the Choptuik scaling relation
\bea
M \simeq C (p-p_{*})^{\gamma},
\label{scaling}
\eea
in the limit $p \sim p_{*}$ with $p > p_{*}$. The constant $\gamma$
is the same for all such one-parameter families, and takes
the numerical value $ \gamma \sim 0.37$, for a four-dimensional
black hole. The parameter $\gamma$ is known as the Choptuik scaling
parameter. Here, $C$ depends on the initial data.

It is of interest to have a model of black hole formation
in which the Choptuik scaling can be calculated in an exact and
analytic fashion.
Our aim here is to show that indeed this can be achieved for the
BTZ black hole.
The key observation is to establish a connection between two
seemingly different spacetime
constructions, the Gott time machine on the one hand, and
the BTZ black hole on the other \cite{BS1}.
We show how the construction of the Gott time machine
leads to a precise formulation
of the order parameter for BTZ black hole
formation.
An exact determination of the corresponding Choptuik
scaling follows immediately.

\subsection{The Gott Time Machine}

In \cite{Gott}, a precise mechanism was presented for the production
of closed timelike  curves.  In particular, the spacetime of two
point particles with mass and boost parameters $\alpha$ and $\xi$,
in $(2+1)$-dimensional spacetime with vanishing cosmological
constant $\Lambda$, was shown to produce closed timelike curves
if the inequality, $\sin \frac{\alpha}{2}\;\cosh \xi > 1$, 
is satisfied. 

In \cite{Carr1, DJH1}, this Gott time machine was analysed
in terms of the group theoretic approach to point particles \cite{DJH2}.
We first note that the $SO(2,1)$ Lorentz group
of Minkowski space is locally equivalent to
$SL(2, {\mathbf R})$. The essential point to recall is
that elements of $SL(2,{\mathbf R})$ are classified according to the value
of their trace. We have
\bea
\mid{\mathrm{Tr}}\; T\mid &<& 2, \;\; 
{\mathrm{Elliptic\; (Timelike)}},\non\\
\mid{\mathrm{Tr}}\; T\mid &=&2, \;\; 
{\mathrm{Parabolic \;(Lightlike)}},\non\\
\mid {\mathrm{Tr}} \;T\mid &>& 2, \;\;
{\mathrm{Hyperbolic \;(Spacelike)}}.
\eea

In the following, we shall use the equivalent $SU(1,1)$ notation
instead of $SL(2,{\mathbf{R}})$;
they are locally equivalent and related by conjugation,
which is given explicitly in \cite{Steif}.
As shown in \cite{DJH2}, the spacetime for a single static point particle 
with $\Lambda = 0$ is obtained by removing a wedge of deficit
angle $\alpha$, and identifying opposite sides
of the wedge. The  particle spacetime is defined via
the rotation generator with angle
$\alpha$,
\bea
R(\alpha) =
\left (
\begin{array}{cc}
e^{-i \alpha/2}&0\\
0& e^{i\alpha/2}
\end{array}
\right ). 
\eea
The mass $m$ of the particle is given by $\alpha = \pi m$, 
in units with $8G=1$, and the resulting spacetime has  a naked conical 
singularity.

A moving particle is obtained by boosting to the rest frame of the 
particle,
rotating, and then boosting back.
Thus, the generator for a moving particle is
\bea
T = B(\mbox{\boldmath $\xi$}) R(\alpha) B^{-1}(\mbox{\boldmath $\xi$}),
\eea
where the boost matrix is given by
\bea
B(\mbox{\boldmath $\xi$}) =
\left (
\begin{array}{cc}
\cosh \frac{\xi}{2}&e^{-i\phi} \sinh \frac{\xi}{2}\\
e^{i\phi} \sinh \frac{\xi}{2}& \cosh \frac{\xi}{2}
\end{array}
\right ).
\eea
Here,  $\mbox{\boldmath $\xi$}$ is the boost vector
with $\xi = |\mbox{\boldmath $ \xi$}|$, and
$\phi$ is the polar angle.

To construct the Gott time machine, we consider a two-body collision
process, with particles labelled by $A$ and $B$. The effective two-particle
generator is then the product $T^{G} = T_{B}T_{A}$
\cite{Carr1}-\cite{DJH2}.
The central object of interest to us is the trace of this generator.
It is straightforward to compute
\bea
\frac{1}{2}\; {\mathrm{Tr}}\;T^{G} &=& \cos \frac{\alpha_{A}}{2}\;\cos
\frac{\alpha_{B}}{2} + \sin \frac{\alpha_{A}}{2}\;\sin\frac{\alpha_{B}}{2}
\non\\
&-& \sin\frac{\alpha_{A}}{2}\;
\sin\frac{\alpha_{B}}{2}\left[ \cosh^{2}\left(\frac{\xi_{A} 
+ \xi_{B}}{2}\right) 
+\cosh^{2}\left(\frac{\xi_{A} - \xi_{B}}{2}\right)\right]\non\\
&+& \sin\frac{\alpha_{A}}{2}\;
\sin\frac{\alpha_{B}}{2}\cos(\phi_{A} - \phi_{B})
\left[ \cosh^{2}\left(\frac{\xi_{A} + \xi_{B}}{2}\right) 
-\cosh^{2}\left(\frac{\xi_{A} - \xi_{B}}{2}\right)\right].
\eea
The original Gott time machine is recovered by choosing particles
with equal masses, and equal and opposite boosts, namely
$\alpha_{A} = \alpha_{B} = \alpha, \xi_{A} = \xi_{B} = \xi, 
\phi_{A} - \phi_{B} = \pi$.
We find  
\bea
\frac{1}{2}\; {\mathrm{Tr}}\;T^{G} = 1 - 2 \sin^{2}\frac{\alpha}{2}\;
\cosh^{2}\xi.
\label{Gott}
\eea
When the Gott condition is satisfied,
we have $\sin^{2}\frac{\alpha}{2}\;\cosh^{2} \xi > 1$, and thus $T^{G}$
is a hyperbolic generator, and when
$\sin^{2} \frac{\alpha}{2} \;\cosh^{2} \xi < 1$, we have
an elliptic generator.
Thus, the effective two-particle generator
(the Gott time machine) becomes hyperbolic (spacelike)
precisely when the Gott condition is satisfied.
The physical consistency of this spacetime has been discussed in
\cite{Carr1, DJH1, Carr2, Menotti}.

\subsection{Choptuik Scaling}

We are interested in applying this algebraic construction to study the
formation of the BTZ black hole. For negative cosmological
constant, the static and moving particle spacetimes are defined
in a fashion analogous to the above, except that one has
both left and right generators. Particle spacetimes for non-zero
cosmological constant have been constructed in \cite{DJ}.
For our purposes here, the most important aspect
of the BTZ black hole is that it is defined by a hyperbolic isometry.
One may choose a fundamental region
of this hyperbolic isometry, and define the black hole spacetime
by identification of the region's boundaries by the isometry.

We recall that the conventional mass parameter of the BTZ
black hole is denoted by $M$, while the point particle mass $m$
is related by 
$m = 2 (1 - \sqrt{-M})$.
As a result, the point particle mass spectrum is $-1 < M < 0$,
while the black hole mass spectrum is $M \geq 0$, with $M=-1$
corresponding to ${\mathrm{AdS}}_{3}$.

Let us consider the static black hole case, in which the left
and right generators are taken to be equal \cite{Carlip1}.
Since the isometries of $AdS_{3}$
are subject to the identification $(\rho_{L}, \rho_{R}) \sim
(-\rho_{L}, -\rho_{R})$, we
may take $\rho_{L} = \rho_{R} = -T^{G} \equiv \rho$.
If the Gott condition is satisfied,
then as we have seen $T^{G}$ is a hyperbolic generator, and consequently
the Gott time machine results in BTZ black hole formation.
The black hole mass is then given by \cite{Carlip1}
\bea
\frac{1}{2}\; {\mathrm{Tr}}\;\rho &=& \cosh \pi \sqrt{M} = -1 + 2 \sin^{2}
\frac{\alpha}{2}\;\cosh^{2}\xi \equiv p,
\label{bhmass}
\eea
where $p \geq 1$.

It is important to consider the defining equation
of this process, namely
\bea
T_{B}T_{A} = T^{G}.
\label{Gott2}
\eea
On the left-hand side, we have the input data given by the particle
mass and boost parameters $\alpha$ and $\xi$. The incoming
particles $A$ and $B$ have been set up in a symmetrical way, with
equal mass and boost parameters, and the timelike geodesics
representing their worldlines will
intersect at a given time, say $t=0$. Thus, we may regard
this as the time of collision of the two particles.
The product of
the particle generators represents the effective generator of the system
at this time \cite{Mats1}. As we have seen, the effective generator
at the time of collision is hyperbolic if the Gott condition is
satisfied. We may then interpret equation (\ref{Gott2})
as defining the formation of a BTZ black hole
at time $t=0$, with the value of the black hole mass
fixed by the input parameters
$\alpha$ and $\xi$. Thus, equation (\ref{Gott2}) encodes
dynamical information.
However, the precise details of the
motion of the particles corresponding to the generators $T_{A}$ and $T_{B}$
prior to the collision, as well as the motion after collision
may also be studied. Indeed, this
analysis has been performed for massless particles in \cite{Mats1,Mats2}.

We see from (\ref{bhmass}) that the natural order parameter
for black hole formation in $(2+1)$-dimensional anti-de Sitter 
gravity is
the trace of the generator. This takes a critical value
at the threshold for black hole formation, corresponding to the 
critical value of the parameter $p_{*} = 1$.
Clearly, $p=p_{*}$ corresponds to the black hole vacuum $M=0$,
where the Gott generator is parabolic. 
Since the parameter $p$ depends on the initial data $\alpha$ and $\xi$,
we can read off the critical boost $\xi$ for any given
mass $\alpha$.
We have 
\bea
\pi \sqrt{M} = {\mathrm{arccosh}}\; p
= \ln\left[p + \sqrt{p^{2} - 1}\right].
\label{exact}
\eea
We stress that the above expression is an exact analytic formula
for the formation of a BTZ black hole
in terms of the input (initial) parameters $\alpha$ and $\xi$,
equivalently $p$. From this, we can immediately
determine the Choptuik scaling by
studying the behaviour near $p_{*}$. 
The mass $M$ and horizon length $r_{+}$ are related by
$\sqrt{M} = r_{+}/l$.
Writing $p = p_{*} + \epsilon$, we find to leading order 
\bea
\frac{r_{+}}{l} = \frac{\sqrt{2}}{\pi} (p - p_{*})^{1/2}.
\eea
Thus, we observe that the horizon length scales with a factor of $1/2$.
Note that this is indeed a universal scaling since BTZ black hole
formation always requires a hyperbolic generator.
The universal scaling value of $1/2$ is simply a consequence
of the fact that the horizon length depends on the inverse cosh function.
One can equally well express the scaling behaviour directly in terms
of the mass, as in (\ref{scaling}).

If the Gott condition is not satisfied, then one has an
effective  particle
spacetime with an elliptic generator, whose effective deficit angle is 
denoted by $\alpha_{\mathrm{eff}}$. This can be obtained by  
continuation of (\ref{bhmass}) to negative $M$ values. We find,
\bea
\frac{1}{2}\; {\mathrm{Tr}}\; \rho &=& \cos  \pi \sqrt{-M} = 
-1 + 2 \sin^{2}\frac{\alpha}{2}\;\cosh^{2}\xi \equiv p,
\eea
where now $p < p_{*}$. 
Once again, we have an exact analytic expression for the 
mass parameter on the other side of the transition, and of course
the Choptuik scaling exponent is again $\gamma = 1/2$, with
\bea
\alpha_{\mathrm{eff}} = 2 \pi - 2 \sqrt{2}(p_{*} - p)^{1/2}.
\eea

One can also consider Choptuik scaling for the spinning black
hole, and in this case we need independent left and right generators.
One finds that $(r_{+} \pm r_{-})$ both have a scaling value
of $1/2$ \cite{BS1}.
The problems of closed timelike curves
and chronology protection \cite{Haw}
are overcome by the creation of the black hole horizon, as 
soon as the Gott condition is satisfied.
We mention that the recent study of BTZ
black hole formation from massless particle collisions \cite{Mats1,Mats2}
is based on the lightlike
analogue of the Gott time machine \cite{DS}, and thus is also
guaranteed to produce the hyperbolic
generator necessary for BTZ formation. 
The holographic description of the creation process has been
investigated within the AdS/CFT
correspondence in \cite{Dan1}-\cite{Louko}.
The formation of the BTZ black hole from gravitational collapse
was first considered in \cite{Mann}, and a Choptuik scaling
value of $1/2$ was also observed for collapsing dust shells
in \cite{Peleg}.
Furthermore, a precise interpretation of this Choptuik scaling
parameter in terms of the timescale for return
to equilibrium of the boundary conformal field theory
has recently been given \cite{DB}.
Numerical studies in $(2+1)$ dimensions
have also been performed recently with different
scaling values. This is explained by the nature of the
transition to the black hole phase \cite{Pret}-\cite{Fabbri}.

\section{Kinematical Properties: Holography and Entropy}

The formulation of a precise holographic principle for
gravitating systems has been widely discussed.
Within the context of string theory, the recent conjecture
due to Maldacena provides a concrete testing ground for such ideas.
In essence, the Maldacena conjecture establishes a duality
between string theory defined on an anti-de Sitter background,
and a corresponding superconformal field theory defined on the
boundary of anti-de Sitter space.
In $2+1$ dimensions, the Maldacena conjecture can be investigated
more thoroughly,
since details are known about both sides of the
AdS/CFT correspondence \cite{DMMV,MS}.
Since solutions to Einstein's equations with
negative cosmological constant
form the basic ingredients in the correspondence,
it is clear that
the BTZ black hole
will play an important role in verifying the Maldacena conjecture.

In this section, we will demonstrate that a precise mathematical
notion of holography can be defined in the context of
three-dimensional hyperbolic geometry. In particular, we
invoke a theorem of Sullivan to establish the exact sense in which the
BTZ black hole is a holographic manifold \cite{BKSW}.
This theorem may also be interpreted as a `No Hair Theorem.'

We then show how the asymptotic symmetry algebra
of $(2+1)$-dimensional gravity (Brown-Henneaux algebra) can be
used to give a microscopic derivation
of the Bekenstein-Hawking entropy for non-extremal BTZ black holes
\cite{Strom, BSS2}.
Furthermore, we employ an exact result from analytic number theory
(a Rademacher expansion) to compute precisely
the corrections to the Bekenstein-Hawking
entropy \cite{BS2}.

\subsection{Sullivan's Theorem}

The theorem of Sullivan \cite{Sull,McM}
provides a precise mathematical statement
of a holographic principle, i.e., a three-dimensional structure
being completely determined in terms of a two-dimensional structure.
Essentially, the theorem states that the inequivalent hyperbolic
structures of a three-dimensional geometrically
finite Kleinian manifold (a term to be defined) are parametrized
by the Teichm\"uller
space of the boundary.
In this section, we explore the consequences of this theorem in the context
of the Euclidean BTZ black hole.

Let us consider the upper half-space model for hyperbolic $3$-space
\cite{Bowditch}, with metric given by (\ref{up}). The boundary at
infinity is the $z=0$ plane and the point $z=\infty$, namely
a $2$-sphere $S^{2}_{\infty}$.
Let $\Gamma$ be a discrete subgroup of isometries of $H^{3}$.
We denote by $\Gamma_{x}$ the orbit of any point of $H^{3}$ under
the action of $\Gamma$.
The {\em limit set} of $\Gamma$ is defined as
\be
L_{\Gamma} = \overline{\Gamma}_{x}\cap S^{2}_{\infty}.
\ee
Thus, the limit set is
the intersection of the closure of $\Gamma_{x}$ with the sphere
at infinity,
for any point $x \in H^{3}$. One can show that the limit set
is independent of the choice of $x$, as we will demonstrate
explicitly in the case at hand.
Given the limit set $L_{\Gamma}$,
one now defines the {\em convex hull}
$H(L_{\Gamma})$ to be the smallest convex set in $H^{3}$
containing $L_{\Gamma}$.
We recall that a convex set in $H^{3}$ is one which contains all
geodesics joining any two points in the set.
The associated {\em convex core}
is obtained as a quotient
\be
C(\Gamma) = H(L_{\Gamma})/\Gamma.
\ee
Thus, the convex core is the quotient by $\Gamma$ of the smallest
convex set in $H^{3}$ containing all geodesics with both end points
in the limit set.

We see that the action of $\Gamma$ partitions $S^{2}_{\infty}$ into
the limit set $L_{\Gamma}$ and its complement $\Omega$, which is
called the {\em domain of discontinuity}.
The resulting Kleinian manifold
\be
N = (H^{3} \cup  \Omega)/\Gamma,
\ee
is then a $3$-manifold with a hyperbolic structure on its interior
and a complex structure on its boundary \cite{Bowditch}.
A hyperbolic $3$-manifold is said to be a {\em geometrically finite}
Kleinian manifold if
the volume of the convex core of $\Gamma$ is finite.
As shown in \cite{Bowditch}, there are several equivalent definitions
of geometrical finiteness.
The main theorem may now be stated as follows.
Let $M$ denote a topological $3$-manifold, and let $GF(M)$ denote
the space of geometrically finite hyperbolic $3$-manifolds $N$
which are homeomorphic to $M$. Thus, $GF(M)$ denotes the space of
realizations of $M$ by geometrically finite Kleinian manifolds $N$.
Then, we have the following theorem due to Sullivan \cite{Sull},
see also \cite{McM}.

\noindent {\bf Theorem}: As long as $M$ admits at least one
hyperbolic realization,
there is a 1-1 correspondence between hyperbolic structures on $M$
and conformal structures on $\partial M$, i.e.,
\be
GF(M) \cong \mathrm{Teich}(\partial M),
\ee
where $\mathrm{Teich}(\partial M)$
is the Teichm\"{u}ller space of $\partial M$.

In order to show that the BTZ black hole
is a geometrically finite Kleinian manifold,
we must first determine the limit
set.
If we let $\gamma$ denote the identification defined by (\ref{Gamma}),
then
clearly the BTZ group is a cyclic Kleinian group with elements
\be
\Gamma_{\mathrm{BTZ}} = \{\gamma^{n}, n \in {\mathbf Z}\}.
\ee
The orbit of any point in $H^{3}$ under the action
of the BTZ group (\ref{Gamma}) is easily obtained in the upper-half space
model.
A simple geometrical picture of the orbit of any point
is obtained by noting that the orbit has a helical structure, whereby
points are rotated around the $z$-axis, as they are translated
upwards or downwards along the $z$-axis.
Such isometries are referred to as
loxodromic, with the $z$-axis called the loxodromic axis \cite{Bowditch}.
Thus, the limit set
of $\Gamma_{\mathrm{BTZ}}$
consists of two points, corresponding to $n\rightarrow
\pm \infty$.
Note that the limit set
is independent of the chosen point, as required.

Kleinian groups which have a finite limit
set are called elementary, and are
known to be geometrically finite \cite{Maskit}.
Since the limit set of $\Gamma_{\mathrm{BTZ}}$ consists of
two points, it follows that $\Gamma_{\mathrm{BTZ}}$ is elementary.
Hence, the  BTZ
black hole is a geometrically finite Kleinian manifold.
Furthermore, as shown in section 2, the topology is that of a
solid torus.

\subsection{Consequences of Sullivan's Theorem}

$\mathbf{\bullet\; No\; Hair\; Theorem}$:
The Euclidean BTZ black hole has the topology of a solid
torus. Since the Teichm\"{u}ller space of the torus
is parametrized by two real parameters,
the theorem can be interpreted as a `No Hair Theorem.'
We conclude that BTZ
black hole can be parametrized by at most
two parameters (mass and angular momentum),
preventing the construction of a charged rotating generalization
as a geometrically finite Kleinian manifold with solid
torus topology.

\noindent
$\mathbf{\bullet\;Holography}$: The theorem
declares that the BTZ black hole
is a holographic manifold, such that
the three-dimensional hyperbolic structure is in 1-1
correspondence with the
Teichm\"{u}ller parameters of the two-dimensional genus one
boundary.

\noindent
$\mathbf{\bullet\; Entropy}$: As the Bekenstein-Hawking
entropy formula is a
geometrical
quantity (the length of the horizon),
it is determined once the hyperbolic structure is fixed.
Hence, the Bekenstein-Hawking entropy is determined
by the Teichm\"{u}ller space on the boundary.

\noindent
$\mathbf{\bullet\;Maldacena\; Conjecture}$: In \cite{MS}, the
action for
the Euclidean BTZ black hole and $H^{3}$ (with periodic time)
were written in
terms of a complex temperature parameter $\tau$ defined on the boundary
torus.
The resulting form of the action then suggested the existence
of an
$SL(2,{\mathbf Z})$ family of solutions whose boundary data $\tau$
is related by the associated modular transformations. We see that the
above theorem does indeed establish the existence of
this class of hyperbolic geometries. Furthermore,
it also follows that two such geometries whose
Teichm\"{u}ller parameters are related by a modular transformation
are equivalent as hyperbolic structures.
In \cite{DMMV}, the Maldacena conjecture was studied in detail for
string theory defined on $AdS_{3} \times S^{3} \times K3$.
The above $SL(2, \mathbf{Z})$ family of solutions
was constructed and played an important role in the investigation.

\subsection{The Bekenstein-Hawking Entropy}

It was observed in \cite{BH} that the asymptotic symmetry algebra of
$(2+1)$-dimensional anti-de Sitter gravity
realises a left-moving and right-moving Virasoro algebra,
with a central charge $c = \overline{c} = 12l$ (in units with
$8G = 1$). To see this, we consider the asymptotic form of the
BTZ metric (\ref{btzmet})
\be
ds^2\to r^2dx^+dx^-+l^{2}\frac{dr^2}{r^2}\ ,
\ee
where $x^\pm=\phi\pm t/l$ are the light-cone coordinates on the
$AdS$-boundary. This asymptotic form is invariant 
under the infinitesimal transformations $x\to x+\xi$ with
\be
\xi^\pm=f^\pm(x^\pm)\ ,\qquad \xi^r=-r\left({f^+}'(x^+)
+{f^-}'(x^-)\right)\ .
\ee
The $\xi^\pm$ generate precisely the group of conformal transformations 
in $1+1$ dimensions. The generators from a representation 
of the Virasoro algebra
\be\label{Vir}
[L_m,L_n]=(m-n)L_{m+n}+\frac{c}{12}(m^3-m)\delta_{m+n,0}\ ,
\ee
and analogously for the left handed generators $\bar L_m$. The central 
charge $c$ is representation dependent. The lowest Fourier components 
$L_0$ and $\bar L_0$ generate global time translations and rotations. 
Comparison with (\ref{boun}) then leads to the identification 
\be
Ml = L_{0} + \bar L_{0},\;\; J = L_{0} - \bar L_{0}\ .
\label{bh}
\ee
Furthermore, it can be shown \cite{BH} that value of the central charge $c$
is a consequence of the ``improvement term"
$\xi^r$, in much the same way
as it appears in constrained WZW models, for example. The physical
interpretation of this result is that, whatever the microscopic theory of 
the BTZ black hole, it will have to form a representation of the conformal
algebra (\ref{Vir}). 

To understand the connection of the algebra (\ref{Vir}) with the black hole 
entropy, we use a result due to Cardy \cite{Cardy}. It relates
the asymptotic degeneracy of states $d(N,\bar N)$ for fixed $L_0=N$ and 
$\bar L_0=\bar N$ to the 
central charge of a unitary, modular invariant conformal field theory by 
\be\label{cardy}
d(N,\bar N)=\exp\left[2\pi\sqrt{\frac{c}{6}N}+2\pi\sqrt{\frac{c}{6}\bar N}\right]\ .
\ee
Let us first consider an extremal black hole $J = Ml$ ($\overline{L}_{0} = 
0$). Taking the logarithm of the degeneracy of states and using (\ref{cardy}), 
(\ref{bh}), and (\ref{rpm}), we find the microcanonical entropy
\be
S = 2 \pi \sqrt{\frac{c}{6}L_{0}} = 4 \pi r_{+}.
\ee
We thus have agreement between the entropy of the
asymptotic conformal field theory and the classical
Bekenstein-Hawking formula. Since the Cardy formula is valid for
large eigenvalues of $L_{0}$ and $\bar L_{0}$, this
derivation holds for very massive black holes.

Furthermore, it was observed by Strominger \cite{Strom}
that in fact the Cardy formula can be used to give a microscopic
derivation of the non-extremal
BTZ black hole, without recourse to string theory or supersymmetry.
The Cardy formula takes the general form
\be
S = 2 \pi \sqrt{\frac{c}{6}L_{0}}
+ 2\pi \sqrt{\frac{c}{6}\bar L_{0}}.
\ee
Using (\ref{bh}), one then finds that
\be
S = 4 \pi r_{+}.
\ee

However, in order for this scenario to be acceptable,
we must have a consistent
theory of quantum gravity on $AdS_{3}$.  One possible realization of 
such a consistent
quantum theory is to embed
the BTZ black hole as a solution of string theory \cite{hyun,BSS2}.
The significance of the BTZ black hole entropy for previous D-brane
computations was also realized \cite{Sfetsos,Strom}. In particular, it
was noticed that the five-dimensional black hole studied
in \cite{SV}
has a near-horizon geometry of the form $AdS_{3} \times S^{2}$.
As a result, one
could give a derivation of the
entropy based solely on the Brown-Henneaux algebra, with
a central charge defined in terms of the D-brane charges.

\subsection{The Rademacher Expansion}
We have seen how the microscopic
degrees of freedom of a conformal field theory
encode information about the entropy of a macroscopic black hole,
via the Cardy formula.
In deriving this formula, the starting point is to consider
the modular invariant partition function of a
unitary conformal field theory defined on a two-torus, namely
\bea
Z(\tau) = \mathrm{Tr}\;e^{2 \pi i(L_{0}
- \frac{c}{24})\tau}.
\eea
Here,  $\tau$ is the modular parameter
and $c$ is the central charge of the conformal field
theory. The Fourier expansion of the partition function
takes the form
\bea
Z(\tau) = \sum_{n\geq 0}F(n) e^{2 \pi i (n-\frac{c}{24})\tau}.
\eea
The black hole entropy in this framework is given by $S = \ln F(n)$,
for large $n$.
To study $S$, we use an
exact convergent expansion, due to Rademacher \cite{Rade},
for the Fourier coefficients of a modular form of weight $\omega$.
This is given by \cite{DMMV}
\bea
F(n) &=& 2 \pi \sum_{m -\frac{c}{24} < 0} \left(\frac{n -\frac{c}{24}}
{|m - \frac{c}{24}|}\right)^{(\omega - 1)/2}F(m) \cdot\non\\
&\cdot & \sum_{k=1}^{\infty} \frac{1}{k}\; Kl\left(n -\frac{c}{24},
m -\frac{c}{24}; k\;\right)
I_{1-\omega}\left(\frac{4 \pi}{k}\sqrt{|m -\frac{c}{24}|
(n -\frac{c}{24})}
\;\right).
\label{Rade}
\eea
Here, $I_{1-\omega}$ is the standard Bessel function and
$Kl(n,m;k)$ is a Kloosterman sum defined by \cite{DMMV},
\bea
Kl(n,m;k) = \sum_{d \in({\mathbf Z}/k{\mathbf Z})^{*}}
\exp\left[ \frac{2 \pi i}{k}(d n + d^{-1}m)\right].
\eea

We are interested in the convergent expansion of $F(n)$ for
$\omega = 0$. For large $n$, this will
correspond to the density of states in the conformal field theory
for large eigenvalues of $L_{0}$.
It is of interest to examine the nature of the leading
terms in the Rademacher expansion.
These are given by setting $m=0$ in (\ref{Rade}), and furthermore
restricting to the $k=1$ term. We find
\bea
F(n) = \frac{(2 \pi)^{3/2}}{12} \;c\;S_{0}^{-3/2}\;e^{S_{0}}\;\left[1-
\frac{3}{8 S_{0}} - \cdots\right],
\label{leading}
\eea
where
\be
S_{0} = 2 \pi\sqrt{\frac{c}{6}\left(n - \frac{c}{24}\right)},
\ee
and we  have used the asymptotic expansion of the Bessel function.
The leading terms in the black hole entropy are then given by
\cite{BS2,Carlip2}
\bea
S = S_{0} - \frac{3}{2} \ln S_{0} + \ln c + \mathrm {constant}.
\label{Cardy}
\eea
This reveals the presence of a logarithmic correction to the
Bekenstein-Hawking entropy $S_{0}$. This can also be understood
as arising from the power-like factor multiplying the
asymptotic density of states \cite{DMMV}.

As an example, for the extremal
BTZ black hole considered above, one finds a logarithmic
correction to the entropy of the form $-3/2 \ln(A/4G)$.
In \cite{Carlip2}, the original derivation of the Cardy formula
was extended to include the first subleading correction. Indeed,
(\ref{Cardy}) is in precise agreement with the formula
derived in \cite{Carlip2}.
One interesting point to note is that the $-3/2\ln S_{0}$ term
first appeared in the quantum geometry formalism \cite{KM, KM2}.
However, the main point to stress here is that
the Rademacher expansion is an
exact convergent expression which determines all
subleading corrections.

\section{Dynamical Properties: Decay Rate}
We have already seen that conformal field theory plays
a crucial role in the kinematical properties of the BTZ black hole.
However, the fact that dynamical
issues such as the decay rate of non-extremal black holes
can also be given a conformal field theory interpretation
is perhaps more surprising.
As already mentioned, following the successful D-brane derivation
of the entropy of extremal black holes in string theory, the properties
of non-extremal black holes were studied. Remarkably, it was
shown that Hawking radiation of near-extremal
black holes
also has an analogue in the  emission of closed strings in D-brane
dynamics;
for a review, see, for example, \cite{Mal1,Horowitz, Peet1, Peet2}.
In the original examples
which were studied \cite{Callan}-\cite{GK},
the calculations required a certain matching
of solutions to the wave equation in an overlap region
between the near-horizon and asymptotic regions.

We begin this section by studying the propagation of a massless
minimally
coupled scalar field in the BTZ background. The key point
is that once again this model allows for an exact calculation,
without the necessity of any matching procedure \cite{Lar,IS}.
Consequently, we can determine precisely the range of energy and
angular momentum
of the scattered field, for which the decay rate
is consistent with
a conformal field theory description \cite{BSS1}.
In particular, we find
that the latter description is not restricted to the
near-extremal limit.
The decay rate for various other fields is presented in
\cite{myung1}-\cite{dasgupta}.
In the second part, we give a microscopic
derivation 
of this semiclassical decay rate by working out the perturbation of the 
conformal field theory realised by the BTZ black hole.

\subsection{The Wave Equation}
It is known that the minimally coupled scalar field
equation can be solved exactly in the background
geometry of the BTZ black hole \cite{Lar,IS}. This allows an
exact determination of
the scattering cross section and decay rate of the scalar field.
The scalar wave equation $\nabla^{2}\Psi = 0$ takes the form
\bea
\left(-f^{-2}\partial_t^2+f^2\partial_r^2+\frac{1}{r}
\left(\partial_r rf^2\right)\partial_r -
\frac{J}{r^2}f^{-2}\partial_t\partial_\phi
-\frac{A}{r^2}f^{-2}\partial_\phi^2\right)
\Psi=0,
\label{box2}
\eea
where
\bea
f^2=\frac{1}{l^2 r^2}(r^2-r_-^2)(r^2-r_+^2),\;\;
A=M-\frac{r^2}{l^2}.
\label{c1}
\eea
Employing the ansatz
\bea
\Psi(r,t,\phi)=R(r,\omega,m)e^{-i\omega t+im\phi},
\label{A1}
\eea
leads to the radial equation for $R(r)$
\bea
\partial_r^2R(r)&+&\left(-\frac{1}{r}+\frac{2r}
{r^2-r_-^2}+\frac{2r}{r^2-r_+^2}\right)\partial_{r}R(r)
+f^{-4}\left(\omega^2-\frac{J\omega m} 
{r^2}+\frac{Am^2}{r^2}\right)R(r).\nonumber\\
\label{box5}
\eea
Upon the change of variables
$z=\frac{r^{2}-r_{+}^{2}}{r^{2} - r_{-}^{2}}$,
the radial equation becomes
\bea
z(1-z)\partial_z^2\;R(z)+(1-z)\partial_z\;R(z)
+\left( \frac{A_{1}}{z} + B_{1}\right) R(z)=0,
\label{box8}
\eea
where  
\bea
A_{1}=\left(\frac{\omega-m\Omega_H}{4\pi T_H}\right)^2,\;\;
B_{1}=-\frac{x_-}{x_+}\left(\frac{\omega
-m\Omega_H\frac{x_+}{x_-}}{4\pi T_H} 
\right)^2.
\label{AB}
\eea

The hypergeometric form of (\ref{box8})
becomes explicit upon removing the pole in the
last term through the ansatz 
\bea
R(z)=z^\alpha g(z),\;\;\alpha^{2} =-A_{1}.
\label{A2}
\eea
We then have
\bea
z(1-z)\partial_z^2g(z)+(2\alpha+1)(1-z)\partial_zg(z)+(A_{1} +B_{1})
g(z)=0.
\label{aldet}
\eea
In the neighbourhood of the horizon, $z=0$,
two linearly independent solutions are then given by 
$F(a,b,c,z)$ and $z^{1-c}F(a-c+1,b-c+1,2-c,z)$, where
\bea
a+b&=&2\alpha,\non\\
ab&=&\alpha^2-B_{1}, \non\\
c&=&1+2\alpha.
\label{abc}
\eea
Note that $c = a+b+1$. 

\subsection{ Semiclassical Decay Rate}
We choose the solution which has ingoing flux at the horizon, namely,
\be
R(z)=z^\alpha F(a,b,c,z).
\label{sol1}
\ee
To see this, we note that the conserved flux for (\ref{box5}) is given,
up to an irrelevant  
normalisation, by  
\bea
{\cal{F}}=\frac{2\pi}{i}\left(R^*\Delta\partial_r R-
R\Delta\partial_r R^*\right),
\label{flux1}
\eea
where $\Delta= r f^2$. The flux can be evaluated by noting that
\bea
\Delta\partial_r=\frac{2\Delta_-}{l^2}z\partial_z,
\label{dz}
\eea
where $\Delta_{-} = r^{2}_{+} - r^{2}_{-}$.
Then, using  the fact that $ab$ is real, 
we find the total flux (which is independent of $z$) to be given by 
\bea
{\cal{F}}(0)=\frac{8\pi\Delta_-}{l^2}{\mathrm Im}[\alpha]
\left|F(a,b,c,0)\right|^2=2{\cal{A}}_H(\omega-m\Omega_H).
\label{f2}
\eea

In order to compute the absorption cross section, we need to
divide (\ref{f2}) by the ingoing flux at infinity.
The distinction between ingoing and outgoing waves is
complicated by the fact
that the BTZ spacetime is not asymptotically flat.
However, we can define ingoing 
and outgoing waves
to be complex linear combinations
of the linearly independent solutions at infinity.
This leads to the definition  \cite{BSS1}
\bea
R_{\mathrm{in}}=A_{\mathrm{in}}
\left(1-i\frac{cl^2}{r^2}\right),\;\; R_{\mathrm{out}}=
A_{\mathrm{out}}\left(1+i\frac{cl^2}{r^2}\right),
\label{inout}
\eea
where $c$ is some positive dimensionless constant, which we take to be 
independent of  the frequency $\omega$, and $R_{\mathrm{in}}$ and
$R_{\mathrm{out}}$ have positive and negative flux, respectively.
The ingoing flux is correspondingly 
\bea
{\cal{F}}_{\mathrm{in}}=8 \pi c|A_{\mathrm{in}}|^2.
\label{fi}
\eea

The asymptotic behaviour of (\ref{sol1}) for large $r$ is readily available
\cite{AS}, and we can then match this to (\ref{inout}) to
determine the coefficients $A_{\mathrm{in}}$ and $A_{\mathrm{out}}$.
We find
\bea
A_{\mathrm{in}}+A_{\mathrm{out}}&=&
\frac{\Gamma(a+b+1)}{\Gamma(a+1)\Gamma(b+1)},\nonumber \\
A_{\mathrm{in}}-A_{\mathrm{out}}&=&-\frac{\Delta_-\Gamma(a+b+1)}{icl^2}
\left\{\frac{\log(\Delta_-/l^2)+\psi(a+1)+\psi(b+1)
-\psi(1)-\psi(2)}{\Gamma(a)\Gamma(b)}\right.\nonumber \\
&+&\left.\frac{\alpha}{\Gamma(a+1)\Gamma(b+1)}\right\},
\label{match} 
\eea
where $\psi$ is the digamma function.
We can estimate the relative importance of the two terms in (\ref{match})
as follows.
If $m=0$ and $\omega<<\hbox{min}(\frac{1}{r_+},\frac{1}{l})$,
the  
difference $A_{\mathrm{in}} - A_{\mathrm{out}}$
in (\ref{match}) is small compared to the
sum $A_{\mathrm{in}} + A_{\mathrm{out}}$, so that
\bea
A_{\mathrm{in}}\sim \frac{1}{2}\frac{\Gamma(a+b+1)}{\Gamma(a+1)
\Gamma(b+1)}.
\label{match4}
\eea
This approximation means that the Compton wavelength of the scattered 
particle is much
bigger that the size of the black hole and the scale set by the curvature
of the anti-de Sitter space.
 
Let us  consider the  
$m =0$ wave and assume
$\omega<<\hbox{min}(\frac{1}{r_+},\frac{1}{l})$
so that (\ref{match4}) is valid. Then the partial wave absorption cross
section  
is given by  
\bea
\sigma^{m=0}=\frac{{\cal{F}}(0)}{\cal{F}_{\mathrm{in}}}=
\frac{1}{\pi c}{\cal{A}}_H
\omega\frac{| 
\Gamma(a+1)\Gamma(b+1)|^2}{|\Gamma(a+b+1)|^2}.
\label{sigma1}
\eea
In order to relate the partial wave cross section to the plane wave cross  
section $\sigma_{abs}$, we need to divide $\sigma^{m=0}$ by $\omega$  
\cite{Gibbons}. We find  
\bea
\sigma_{abs}= {\cal{A}}_H\frac{|
\Gamma(a+1)\Gamma(b+1)|^2}{|\Gamma(a+b+1)|^2},
\label{sigma2}
\eea
where we have chosen $c$ so that $\sigma_{abs}(\omega)
\rightarrow{\cal{A}}_H $  
for $\omega\rightarrow 0$ \cite{Gibbons}. The decay rate $\Gamma$ of a 
non-extremal black hole is then given by \cite{BSS1}
\bea
\Gamma&=&\frac{\sigma_{abs}}{e^{\frac{\omega}{T_H}}-1}=
T_H{\cal{A}}_H\omega^{-1}e^{-\frac{\omega}{2T_H}}| 
\Gamma(a+1)\Gamma(b+1)|^2\non\\
&=&4\pi^2l^2\omega^{-1}T_LT_Re^{-\frac{\omega}{2T_H}}\left|
\Gamma\left(1+i\frac{\omega}{4\pi T_L}\right)\Gamma\left(1+i
\frac{\omega}{4\pi T_R}\right)\right|^2,
\label{decay1}
\eea
where the left and right temperatures are defined by  
\bea
T_{L/R}^{-1}=T_H^{-1}\left(1\pm\frac{r_-}{r_+}\right),
\label{TLR}
\eea
and we have used ${\cal{A}}_H=2\pi r_+$.

\subsection{Microscopic Description}
In this subsection, we reverse the usual rules for computing the
Hawking decay rate. That is we quantize the black hole while considering 
the matter field of the last subsection as an external field \cite{ES}. 
Matter fields perturb 
the dynamics of the metric by acting as sources of 
energy and momentum. The field $\Psi$ is treated classically, i.e. taken 
to satisfy the classical wave equation in the bulk 
of 
$AdS_3$. One may think of this as the curved space equivalent of taking a 
homogeneous external field in the case of the atom in a radiation field. 
In this approximation one does not resolve the detailed structure of the 
bulk. The matter action then reduces to a boundary term
\bea
S_{m}= -\frac{1}{2}\int\sqrt{-g}g^{\mu\nu}\partial_\mu
\Psi\partial_\nu\Psi 
\rightarrow -\frac{1}{2}{\cal{B}}^r(\infty),
\label{S_m}
\eea
where
\bea
{\cal{B}}=\frac{1}{2} \int\limits_{\partial {\cal{M}}}\sqrt{-g}g^{rr}
(\partial_r\Psi^\dagger\Psi+ \Psi^\dagger\partial_r\Psi)
\eea
denotes the boundary term. The operator $\sqrt{-g}g^{rr}
\partial_r$ acting on the external field $\Psi$ is the analogue of the
dipole 
operator coupling to an external electric field. How this operator is
represented in the microscopic theory of the black hole will, of course, 
depend on 
the concrete microscopic model for the black hole (see \cite{ES} for 
details). However, for the present purpose it is enough to note that,
according to the results of Brown-Henneaux \cite{BH}, this operator
transforms as a $(1,1)$ primary field under the asymptotic conformal 
isometries of $AdS_3$. Thus,
the external field introduces a 
perturbation of the CFT at the boundary at infinity by a primary 
operator ${\cal O}(\sigma^+,{\sigma^-})$ with conformal weight $(1,1)$. 
Here, $ \sigma^\pm$ denote the light cone coordinates on the asymptotic 
boundary of the black hole metric (\ref{btzmet}). 

Having found the coupling of the external field to the intrinsic degrees 
of 
freedom of the black hole, we can now
compute transition amplitudes occurring 
in the presence of a matter field. As explained above, this
interaction vertex should correctly describe the transition 
between black hole states with small energy difference. 
Note that it is not required that the initial state itself 
has low energy. 

The transition amplitude between an initial and a final state in 
the presence of an external flux with frequency and angular momentum 
$\omega, m$ 
is then given by 
\be\label{tr1}
{\cal M} = \ell\int\;d {\sigma^+} d {\sigma^-}\;<\!f|{\cal 
O}({\sigma^+},{\sigma^-})|i>\!
\;e^{-i(\omega\ell-m){{\sigma^+}\over 
2}}\;e^{-i(\omega\ell+m){{\sigma^-}\over 2}},
\ee
where $i$ and $f$ denote the initial and final black hole state,
respectively. The important point is that calculation 
of transition amplitudes is reduced to the computation of correlation 
functions of $(1,1)$ primary fields. 

We proceed to compute the decay rate. For simplicity we set $m=0$. 
Squaring the amplitude 
(\ref{tr1}) and 
summing over final states leads to 
\be
\sum\limits_f|{\cal{M}}|^2 = \ell^2\int\;d {\sigma^+}d {\sigma^+}'d 
{\sigma^-}d 
{\sigma^-}'\;\;<\!i|{\cal O}({\sigma^+},{\sigma^-})
{\cal O}({\sigma^+}',{\sigma^-}')|i>\!\;e^{-i 
\omega\ell{{\sigma^+}-{\sigma^+}'\over 2}}
\;e^{-i \omega\ell{{\sigma^-}-{\sigma^-}'\over 2}}.
\ee
Since the black hole corresponds to a thermal state, we must average 
over initial states weighted by the Boltzmann factor. This means that we
must 
take finite temperature two point functions, which for fields of conformal 
weight one are given by 
\be
<\!{\cal O}(0,0)
{\cal O}({\sigma^+},{\sigma^-})>\!_{T_R,T_L} =
\left[\frac{\pi T_R}{\sinh(\pi T_R {\sigma^+})}\right]^2
\left[\frac{\pi T_L}{\sinh(\pi T_L {\sigma^-})}\right]^2,
\ee
provided $T>\!>V^{-1}$. 
These have the right periodicity properties in the Euclidean section.
The remaining integrals can be performed by contour techniques of common 
use 
in thermal field theory.  
Whether we deal with emission or absorption depends on how the poles at 
${\sigma^+}=0$, ${\sigma^-}=0$ are dealt with. The choice for emission 
leads to integrals of 
the type
\be
\int d{\sigma^+}\; {e^{-{i\omega\over 2} ({\sigma^+}-i\epsilon)} \over 
\sinh^2(xu)} 
={\pi\omega\over x^2}\sum_{n=1}^\infty e^{-{\omega\pi n\over 2x}}= 
{\pi\omega\over x^2}\left( e^{\omega\pi \over 2x}-1\right)^{-1}\, .
\ee
The resulting emission rate is then given by 
\be\label{decay}
\Gamma = {\omega \pi^2 \ell^2 \over (e^{\omega\over 2T_L} -1) 
(e^{\omega\over2T_R} 
-1)},
\ee
where we have included a factor $\omega^{-1}$ for the normalization of the 
outgoing scalar. Eq.~(\ref{decay}) reproduces correctly the semiclassical 
result (\ref{decay1}), therefore providing a microscopic derivation
of the decay of BTZ black holes. Note that the above analysis is valid for 
all values of $M$ and $J$, 
subject to the low energy restriction.  
Thus, the conformal field theory description of the BTZ black hole is
not restricted to the near-extreme (near BPS) limit.
As in the case of the entropy, the relevance of the above
calculation to the five-dimensional black hole should also be noted.
In particular, the non-trivial part of the greybody factors
of the five-dimensional black hole arises
due to the presence of the BTZ black hole
in the near-horizon limit \cite{I}.
Other aspects of the decay rate within the context of the
AdS/CFT correspondence
are treated in \cite{satoh}-\cite{teo}.

\section{Outlook}
To summarise, we have reviewed various exact results
for the BTZ black hole in $2+1$ dimensions.
In particular, we have shown that this toy model provides
the first example of an exact determination of the Choptuik scaling
parameter. We have also seen that the BTZ black hole provides a precise mathematical
model of a holographic manifold. Furthermore, we found that
the notion of holography (in the sense Sullivan's theorem)
is equivalent to a `No Hair Theorem' for this black hole spacetime.
We found that geometrical properties of the spacetime (as encoded
is the Brown-Henneaux asymptotic symmetry algebra) are sufficient
to yield a microscopic understanding of the Bekenstein-Hawking
entropy. 

In spite of considerable progress,
there are a number of issues that deserve further consideration.
For example, one would like to
construct spacetimes with the topology of a genus $g$ handlebody
\cite{beng1}-\cite{krasnov}.
Indeed, such spacetimes can be constructed via a Schottky
procedure, and are known to be geometrically finite \cite{Maskit}.
Thus,
according to Sullivan's theorem the resulting spacetimes
would be parametrized
by $(3g-3)$ complex Teichm\"{u}ller parameters.
The construction of such spacetimes will play a role
in verifying the AdS/CFT correspondence
on higher genus
Riemann surfaces \cite{krasnov}.

An important issue in any discussion of black hole physics
is the information puzzle. Given the
detailed understanding of
various microscopic aspects of the BTZ black hole,
one might expect that one should be able to gain a better understanding of
the information puzzle in this model. However, a clear resolution of the 
puzzle remains elusive to date. In particular, it would be useful to 
understand the precise relation between the counting of microstates 
presented here and Carlip's counting of states at the horizon 
\cite{Carlip1b}.
\vspace*{1cm}

\noindent{\large \bf Acknowledgements}\\
D.B. and S.S. would like to thank Conall Kennedy and Andy Wilkins
for collaboration in the results of Ref.$\;$\cite{BKSW}. I.S. would 
like to thank Roberto Emparan for collaboration in the results of Ref.$\;$
\cite{ES}. The work of D.B. and I.S. is part of a project supported
by Enterprise Ireland international collaboration grant IC/2000/009.


\begin{thebibliography}{999}
\bibitem{SV} A. Strominger and C. Vafa, Phys. Lett. B 379 (1996) 99;
hep-th/9601029.
\bibitem{Mal1} J. Maldacena, {\em Black Holes in String Theory},
hep-th/9603135.
\bibitem{Horowitz} G. Horowitz, {\em Quantum States of Black Holes},
gr-qc/9704072.
\bibitem{Peet1} A.W. Peet, Class. Quantum Grav. 15 (1998) 3291;
hep-th/9712253.
\bibitem{Skend} K. Skenderis, {\em Black Holes and Branes in String 
Theory},
hep-th/9901050.
\bibitem{Peet2} A.W. Peet, {\em TASI Lectures on Black Holes
in String Theory}, hep-th/0008241.
\bibitem{Pol} J. Polchinski, Phys. Rev. Lett. 75
(1995) 4724; hep-th/9510017.
\bibitem{BTZ1} M. Ba\~{n}ados, C. Teitelboim, and J. Zanelli,
Phys. Rev. Lett. 69 (1992) 1849; hep-th/9204099.
\bibitem{Carlip1} S. Carlip,
Class. Quantum Grav. 12 (1995) 2853; gr-qc/9506079.
\bibitem{BTZ2} M. Ba\~{n}ados, M.Henneaux, C. Teitelboim, and
J. Zanelli, Phys. Rev. D 48 (1993) 1506; gr-qc/9302012.
\bibitem{hyun} S. Hyun, {\em $U$-duality between Three and Higher
Dimensional Black Holes}, hep-th/9704005.
\bibitem{Sfetsos}  K. Sfetsos and K. Skenderis, Nucl. Phys. B 517 (1998)
179; hep-th/9711138.
\bibitem{Strom} A. Strominger, J. High Energy Phys. 02 (1998) 009; 
hep-th/9712251.
\bibitem{Mats1} H.-J. Matschull, Class. Quantum Grav. 16 (1999) 1069;
gr-qc/9809087.
\bibitem{Mats2} S. Holst and H.-J. Matschull, Class. Quantum Grav. 16
(1999) 3095; gr-qc/9905030.
\bibitem{BS1} D. Birmingham and S. Sen, Phys. Rev. Lett. 84 (2000) 1074;
hep-th/9908150.
\bibitem{Chop} M.W. Choptuik, Phys. Rev. Lett. 70 (1993) 9.
\bibitem{Gund} C. Gundlach, Adv. Theor. Math. Phys. 2 (1998) 1;
gr-qc/9712084.
\bibitem{Sull} D. Sullivan,
in {\em Proceedings of the 1978 Stony Brook Conference on Riemann Surfaces
and Related Topics}, edited by I. Kra and B. Maskit,
Annals of Mathematics Studies No. 97, Princeton University Press,
Princeton, New Jersey, 1981.
\bibitem{McM} C. McMullen, Bull. Am. Math. Soc. 27 (1992) 207;
Invent. Math. 99 (1990) 425.
\bibitem{BKSW} D. Birmingham, C. Kennedy, S. Sen, and A. Wilkins,
Phys. Rev. Lett. 82 (1999) 4164; hep-th/9812206.
\bibitem{thooft} G. `t Hooft, {\em Dimensional Reduction in Quantum 
Gravity}, gr-qc/9310026.
\bibitem{Suss} L. Susskind, L. Thorlacius, and J. Uglum, Phys. Rev. D 28
(1993) 3743; hep-th/9306069.
\bibitem{Malda1} J. Maldacena, Adv. Theor. Math. Phys.
2 (1998) 231; hep-th/9711200.
\bibitem{Poly} S.S. Gubser, I.R. Klebanov and A.M. Polyakov,
Phys. Lett. B 428 (1998) 105; hep-th/9802109.
\bibitem{Witten1} E. Witten,
Adv. Theor. Math. Phys. 2 (1998) 253; hep-th/9802150.
\bibitem{Witten2} E. Witten,
Adv. Theor. Math. Phys. 2 (1998) 505; hep-th/9803131.
\bibitem{Malda2} O. Aharony, S.S. Gubser, J. Maldacena, H. Ooguri,
and Y. Oz, Phys. Rep. 323 (2000) 183; hep-th/9905111.
\bibitem{BH} J.D. Brown and M. Henneaux,
Commun. Math. Phys. 104 (1986) 207.
\bibitem{Cardy} J.L. Cardy, Nucl. Phys. B 270 (1986) 186.
\bibitem{BSS2} D. Birmingham, I. Sachs, and S. Sen,
Phys. Lett. B 424 (1998) 275; hep-th/9801019.
\bibitem{BL} V. Balasubramanian and F. Larsen, Nucl. Phys. B 528 (1998)
229; hep-th/9802198.
\bibitem{CL1} M. Cveti\v{c} and F. Larsen, Nucl. Phys. B 531 (1998)
239; hep-th/9805097.
\bibitem{CL2} M. Cveti\v{c} and F. Larsen, Phys. Rev. Lett. 82 (1999)
484; hep-th/9805146.
\bibitem{BS2} D. Birmingham and S. Sen, {\em Exact Black Hole
Entropy Bound in Conformal Field Theory}, hep-th/0008051, Phys. Rev. D
(to appear).
\bibitem{Rade} H. Rademacher, {\em Topics
in Analytic Number Theory}, Springer-Verlag, Berlin, 1973.
\bibitem{DMMV} R. Dijkgraaf, J. Maldacena, G. Moore, and E. Verlinde,
{\em A Black Hole Farey Tail}, hep-th/0005003.
\bibitem{Carlip2} S. Carlip, Class. Quantum Grav. 17 (2000) 4175;
gr-qc/0005017.
\bibitem{Carlip1b} S. Carlip, Phys. Rev. D 51 (1995) 632; gr-qc/9409052.
\bibitem{Callan} C. Callan and J. Maldacena,
Nucl. Phys. B 472 (1996) 591; hep-th/9602043.
\bibitem{mathur} S. Das and S. Mathur,
Nucl. Phys. B 478 (1996) 561; hep-th/9606185.
\bibitem{StromMald1} J. Maldacena and A. Strominger,
Phys. Rev. D 55 (1997) 861;
hep-th/9609026.
\bibitem{StromMald2} J. Maldacena and A. Strominger,
Phys. Rev. D 56 (1997) 4975; hep-th/9702015.
\bibitem{CL} M. Cveti\v{c} and F. Larsen,
Phys. Rev. D 56 (1997) 4994, hep-th/9705192;
Nucl. Phys. B 506 (1997) 107, hep-th/9706071.
\bibitem{GK} S.S. Gubser and I. Klebanov,
Phys. Rev. Lett. 77 (1996) 4491;
hep-th/9609076.
\bibitem{Lar} K. Ghoroku and A.L. Larsen, Phys. Lett. B 328 (1994) 28.
\bibitem{IS} I. Ichinose and Y. Satoh,
Nucl. Phys. B 447 (1995) 340;
hep-th/9412144.
\bibitem{BSS1} D. Birmingham, I. Sachs, and S. Sen,
Phys. Lett. B 413 (1997) 281; hep-th/9707188.
\bibitem{ES} R. Emparan and I. Sachs, Phys. Rev. Lett. 81 (1998) 2408; 
hep-th/9806122.
\bibitem{Carlip3} S. Carlip and C. Teitelboim, Phys. Rev. D 51 (1995) 622;
gr-qc/9405070.
\bibitem{Gott} J.R. Gott, Phys. Rev. Lett. 66 (1991) 1126. 
\bibitem{Carr1} S.M. Carroll, E. Farhi, and A.H. Guth, 
Phys. Rev. Lett. 68 (1992) 263.  
\bibitem{DJH1} S. Deser, R. Jackiw, and G. 't Hooft, 
Phys. Rev. Lett. 68 (1992) 267.
\bibitem{DJH2} S. Deser, R. Jackiw, and G. 't Hooft, Ann. Phys. (N.Y.) 152
(1984) 220.
\bibitem{Steif} A.R. Steif, Phys. Rev. D 53 (1996) 5527; gr-qc/9511053.
\bibitem{Carr2} S.M. Carroll, E. Farhi, A.H. Guth, and K.D. Olum,
Phys. Rev. D 50 (1994) 6190; gr-qc/9404065.
\bibitem{Menotti} P. Menotti and D. Seminara, Ann. Phys. (N.Y.)
240 (1995) 203; gr-qc/9406016. 
\bibitem{DJ} S. Deser and  R. Jackiw, Ann. Phys. (N.Y.) 153
(1984) 405.
\bibitem{Haw} S.W. Hawking, Phys. Rev. D 46 (1992) 603.
\bibitem{DS} S. Deser and A.R. Steif, Class. Quantum Grav. 9 (1992) 
L153; hep-th/9208018. 
\bibitem{Dan1} U.H. Danielsson, E. Keski-Vakkuri, and M. Kruczenski,
Nucl. Phys. B 563 (1999) 279; hep-th/9905227.
\bibitem{Bal} V. Balasubramanian and S.F. Ross, Phys. Rev. D 61 (2000)
044007; hep-th/9906226.
\bibitem{Dan2} U.H. Danielsson, E. Keski-Vakkuri, and M. Kruczenski,
JHEP 02 (2000) 039; hep-th/9912209.
\bibitem{Louko} J. Louko, D. Marolf, and S.F. Ross, Phys. Rev. D 62
(2000) 044041; hep-th/0002111.
\bibitem{Mann} S.F. Ross and R.B. Mann, Phys. Rev. D 47 (1993) 3319;
hep-th/9208036.
\bibitem{Peleg} Y. Peleg and A.R. Steif, Phys. Rev. D 51 (1995) R3992;
gr-qc/9412023. 
\bibitem{DB} D. Birmingham, {\em Choptuik Scaling and Quasinormal
Modes in the AdS/CFT Correspondence}, hep-th/0101194.
\bibitem{Pret} F. Pretorius and M.W. Choptuik, Phys. Rev. D 62 (2000)
124012; gr-qc/0007008.
\bibitem{gar} D. Garfinkle, Phys. Rev. D 63 (2001) 044007; gr-qc/0008023.
\bibitem{hus} V. Husain and M. Olivier, Class. Quantum Grav. 18 (2001)
L1; gr-qc/0008060.
\bibitem{Fabbri} G. Cl\'{e}ment and A. Fabbri, {\em Analytical Treatment
of Critical Collapse in 2+1 dimensional AdS Spacetime}, gr-qc/0101073.
\bibitem{MS} J. Maldacena and A. Strominger, JHEP 9812 (1998) 005;
hep-th/9804085.
\bibitem{Bowditch} B.H. Bowditch,
J. Funct. Anal. 113 (1993) 245.
\bibitem{Maskit} B. Maskit, {\em Kleinian Groups}, Springer-Verlag,
Berlin, 1988.
\bibitem{KM} R.K. Kaul and P. Majumdar, Phys. Rev. Lett. 84 (2000)
5255; gr-qc/0002040.
\bibitem{KM2} R.K. Kaul and P. Majumdar, Phys. Lett. B 439 (1998) 267;
gr-qc/9801080.
\bibitem{myung1} H.W. Lee, N.J. Kim, and Y.S. Myung, Phys. Lett.
B 441 (1998) 83; hep-th/9803227.
\bibitem{myung2} H.W. Lee and Y.S. Myung, Phys. Rev. D 58 (1998)
104013; hep-th/9804095.
\bibitem{dasgupta} A. Dasgupta, Phys. Lett. B 445 (1999) 279;
hep-th/9808086.
\bibitem{AS} M. Abramowitz and I.A. Stegun, {\em Handbook of
Mathematical Functions}, Dover, New York,  1970.
\bibitem{Gibbons} S. Das, G. Gibbons, and S. Mathur,
Phys. Rev. Lett. 78 (1997) 417; hep-th/9609052.
\bibitem{I} I. Sachs, {\it in} PASCOS 98, p. 670, {\it World Scientific}
(1999); hep-th/9804173.
\bibitem{satoh} Y. Satoh, Phys. Rev. D 59 (1999) 084010; hep-th/9810135.
\bibitem{das} S. Das and A. Dasgupta, JHEP 9910 (1999) 025; hep-th/9907116.
\bibitem{ohta} H.J.W. Muller-Kirsten, N. Ohta, and J.-G. Zhou,
Phys. Lett. B 445 (1999) 287; hep-th/9809123.
\bibitem{teo} E. Teo, Phys. Lett. B 436 (1998) 269; hep-th/9805014.
\bibitem{beng1} S. \AA minneborg, I. Bengtsson, D. Brill,
S. Holst, P. Peld\'{a}n, Class. Quantum Grav. 15 (1998) 627.
\bibitem{beng2} S. \AA minneborg, I. Bengtsson, and S. Holst,
Class. Quantum. Grav. 16 (1999) 363.
\bibitem{brill} D. Brill, {\em Black Holes and Wormholes
in $2+1$ Dimensions}, gr-qc/9904083.
\bibitem{krasnov} K. Krasnov, {\em Holography and Riemann Surfaces},
hep-th/0005106.
\end{thebibliography}
\end{document}